# Routine Hospital-based SARS-CoV-2 Testing Outperforms State-based Data in Predicting Clinical Burden

Len Covello, Community Hospital, Munster, Indiana

Andrew Gelman, Departments of Statistics and Political Science, Columbia University, New York, NY

Yajuan Si, Institute for Social Research, University of Michigan, Ann Arbor, MI

Siquan Wang, Department of Biostatistics, Columbia University, New York, NY

Corresponding author: Yajuan Si, Email: yajuan@umich.edu, Telephone: 734-7646935, Survey Research Center, Institute for Social Research, University of Michigan, ISR 4014, 426 Thompson St, Ann Arbor, MI 40104

Running head: A Proxy Metric for Community Coronavirus Tracking

Conflict of interest: none declared.

Funding support: This study was supported by the Michigan Institute of Data Science, National Science Foundation and National Institutes of Health.

**Reproducibility:** Codes are publicly available at https://github.com/yajuansi-sophie/covid19-mrp. The data are confidential and cannot be released to the public.

# Routine Hospital-based SARS-CoV-2 Testing Outperforms State-based Data in Predicting Clinical Burden

ABSTRACT

Throughout the COVID-19 pandemic, government policy and healthcare implementation responses have been guided by reported positivity rates and counts of positive cases in the community. The selection bias of these data calls into question their validity as measures of the actual viral incidence in the community and as predictors of clinical burden. In the absence of any successful public or academic campaign for comprehensive or random testing, we have developed a proxy method for synthetic random sampling, based on viral RNA testing of patients who present for elective procedures within a hospital system. We present here an approach under multilevel regression and poststratification (MRP) to collecting and analyzing data on viral exposure among patients in a hospital system and performing statistical adjustment that has been made publicly available to estimate true viral incidence and trends in the community. We apply our MRP method to track viral behavior in a mixed urban-suburban-rural setting in Indiana. This method can be easily implemented in a wide variety of hospital settings. Finally, we provide evidence that this model predicts the clinical burden of SARS-CoV-2 earlier and more accurately than currently accepted metrics.

**Word counts:** Abstract (188); main text (3999).

INTRODUCTION

Early knowledge of incidence and trends of SARS-Cov-2 transmission in communities is crucial, but in the absence of universal screening or random testing, interested parties have been left to extrapolate impressions of community viral behavior from nonrepresentative data. Public health professionals have relied on state-sourced positivity rates and raw numbers of positive tests in any given jurisdiction as proxies for the true SARS-Cov-2 burden. Unfortunately, these presumed proxies are subject to significant selection bias, as most testing protocols understandably target symptomatic and presumed-exposed populations. Further, tests have been applied with different criteria over time and geography according to test availability, perceived community SARS-Cov-2 burden, and disparate clinical or political testing norms. The uncontrolled nature of the data raises questions or criticism about their validity as determinants of policy that delimits clinical or economic behavior. Alleva et al. provide a review of strategies and experiences currently in progress to estimate the SARS-CoV-2 incidence in the community (1). Briefly, the existing approaches include massive test campaigns without a formal sampling design, diagnostic tests through a probabilistic sample, volunteer massive surveys and supplements of existing sample surveys. Absent randomized testing of the population or community, we need a means of normalizing currently available data to better track trends in the true underlying incidence, either as a more reliable metric or as a reassurance of the validity of our current ones in predicting clinical behavior of SARS-CoV-2.

In the present article, we apply multilevel regression and poststratification (MRP), a standard adjustment method used in survey research that is particularly effective when sample sizes are small in some demographic or geographic slices of the data (2, 3). MRP has increasingly been shown to be useful in public health surveys, and even be successful in highly unrepresentative probability or non-probability samples (4-6). We work with data from the Community Hospital group in Indiana, which serves an urban-suburban-rural mix of patients. COVID testing was already being performed for patients in this hospital system, and it was relatively costless to augment this data collection with the statistical analysis presented here. For this reason, we believe that this method can be easily implemented in a wide variety of hospital settings.

METHODS

Study Data and Sample

Upon reopening without restriction to elective medical and surgical procedures after the early spring COVID-19 outbreak, clinical professionals in our hospital system were sufficiently concerned about asymptomatic viral shedding to test all patients for acute SARS-Cov-2 infection before performing any such procedure. All elective patients for invasive procedures are presumptively asymptomatic, as any potential surgical patient acknowledging symptoms or presenting a recent history of known viral exposure would have the procedure canceled or deferred. All prospective surgical (and other invasive procedure) patients were subjected to a preoperative evaluation of these issues and excluded if they showed evidence of symptoms or exposure.

This population presented a potentially valuable resource. All patients used the same test administered within a health system by similar health care professionals. There is a broad age, racial/ethnic, and economic diversity to this group, and its only overt correlation to disease status is that it is specifically selected for a lack of symptoms and a negative exposure history. By way of contrast, cumulative state-wide data include testing data from multiple sources (private clinics, large hospitals, pop-up clinics, large employers, universities, etc.), with different types of tests, different levels of training for testers, different test settings (clinic office or drive through clinic), where the test results are valuable but likely much more variable than those under consideration. More importantly, the criteria used in the cumulative state data to determine whether to test individuals in the first place are subject to varying prior assumptions. As an example, an outpatient exposed to a family member suspected of active COVID has a substantially different prior than a dyspneic patient admitted to the emergency department, yet the implication of the state data trends is that these positive cases should be handled similarly.

Though not ideal, our test group is therefore a promising proxy for the general community. SARS-CoV-2 has clearly shown the ability to spread throughout the population via both asymptomatic and symptomatic infection. If we were to assume the as yet unverified but reasonable hypothesis that, for any uniform demographic, the ratio of asymptomatic-to-symptomatic viral infection is constant, then the asymptomatic population in a community would vary in a strict ratio with overall prevalence, and could therefore serve as an excellent proxy for true viral incidence. The trending of asymptomatic infection would be expected to be strictly proportional to clinical infection. To the degree to which external comorbidities or other factors might contribute to variation in this ratio across the sample, we anticipate that much of this variation would be captured by our demographic adjustments.

Of crucial importance, our sample group varies from a true random sample in predictable ways. It is selected rigorously for asymptomatic/non-exposed status, and age, racial/ethnic, and geographic demographics are well documented in the hospital electronic health records (EHR). It remains only to normalize our sample to the demographics of the larger community to represent the general population.

Measures

We subjected all patients to polymerase chain reaction (PCR) testing for viral RNA, 4 days before their intended procedure. Samples were submitted to LabCorp for analysis using the Roche cobas system. This testing regime was used throughout the study interval and continues to be employed without change to the present day. A 70% clinical sensitivity is presumed for this test, based on near 100% internal agreement with positive controls on in vitro analytics (7) and broadly observed clinical performance of PCR testing throughout the pandemic (8); however, asymptomatic and pre-symptomatic patients may be harder to detect than predicted by these analytic data, as dates of infection as well as symptom status/onset are known to have a large effect on sensitivity (9). These effects would need to be acknowledged and, to the degree possible, accounted for in the model. Specificity is near 100%, with false positives likely generated only by cross-contamination or switched samples. These false positives become important when underlying prevalence is near zero (10), as was the case for our community this summer, and we have applied a Bayesian procedure to account for the false positivity. We evaluate whether the estimated trends and magnitudes are robust against the sensitivity and specificity parameters.

Statistical Analysis

We are interested in rates of SARS-CoV-2 infection in two populations: 1) Individuals undergoing care within the hospital system as patients, and 2) the community from which the hospital draws as a whole. In addition to adjusting for measurement error associated with PCR testing for SARS-CoV-2 infection, we need to generate standardized estimates that reflect prevalence in the populations of interest rather than merely our sample of elective surgery patients we are drawing on. We anticipate that this sample of asymptomatic patients is a fairly representative group with minor discrepancy selected from the community-at-large, but also expect that poststratification to the target population with matching sociodemographics would help enhance the accuracy of our conclusions. We acknowledge that those who seek even elective hospital-based procedures may further vary from the overall community with respect to their comorbidities (hence, their susceptibility to symptomatic COVID-19), but believe it is reasonable to infer that MRP normalization of their age, race/ethnicity, gender, and geography will account for much of that discrepancy.

We use a Bayesian approach to account for unknown sensitivity and specificity and apply MRP to testing records for population representation, here using the following adjustment variables: reported sex, age (0-17, 18-34, 35-64, 65-74, and 75+), race (white, black, and other), and county (Lake and Porter). MRP has two key steps: (1) fit a multilevel model for the prevalence with the adjustment variables based on the testing data; and (2) poststratify using the population distribution of the adjustment variables, yielding prevalence estimates in the target population.

We poststratify to two different populations: patients in the hospital database (those who have historically and currently obtained care in our regional hospital system) and residents of Lake/Porter County, Indiana. For the hospital, we use the EHR database to represent the population of patients from three hospitals in the Community Health System (Community Hospital, St. Catherine Hospital, and St. Mary Medical Center). For the community, we use the American Community Survey 2014-2018 data from the two counties.

We particularly care about changes in SARS-Cov-2 incidence over time. Indeed, even if our demographic and geographic adjustment is suspect (given systematic differences between sample and populations), the greatest clinical utility lies in being able to predict how much the clinical burden present today is likely to change in the future. Here, the adjustment may be particularly important, as the mix of patients has changed somewhat during the study interval. The statistical details are included in the Supplement. We perform all computations in R (11); data and code are publicly available at https://github.com/yajuansi-sophie/covid19-mrp.

Assumptions and Conjectures

We began the data collection with a few hypotheses or speculations. First, we expected that the ratio between asymptomatic and symptomatic patients would be relatively constant, for a uniform demographic distribution specific to age, gender, and race/ethnicity. Second, we anticipated that changes in PCR positivity among asymptomatic individuals would precede changes in symptomatic PCR-detected infections by several days, because of the known temporal relationship of viral shedding to the onset of clinical disease (12). Third, these hypotheses would imply that trends in our asymptomatic SARS-COV-2 infections would predict the behavior of the virus within the community as a whole. To this end, we aim to determine whether our model mirrors or predicts hospitalization rates as a proxy for clinical viral burden.

In summary, we anticipated that appropriate modeling of the PCR dataset would allow us to measure changes in acute infection incidence as an early warning metric to grasp the developing trend of the disease, or at least in concert with any changes. The procedure provides accurate assessment of trends, rather than incidence, and offers more temporally relevant information than the current use of percent-testing-positive. Further, we aimed to evaluate the validity of positivity and counts of positive cases as metrics to predict clinical burden.

RESULTS

Demographic Stability

Table 1. Descriptive summary of test results and sociodemographic distributions.

|  | Asymptomatic PCR | Symptomatic PCR | Hospital | Community |
|---|---|---|---|---|
| Size | 30116 | 13960 | 35838 | 654890 |
| Prevalence(%) | 1 | 26 | NA | NA |
| Female(%) | 59 | 60 | 57 | 51 |
| Male(%) | 41 | 40 | 43 | 49 |
| Age0-17(%) | 3 | 15 | 9 | 24 |
| Age18-34(%) | 10 | 20 | 12 | 21 |
| Age35-64(%) | 46 | 44 | 30 | 40 |
| Age65-74(%) | 24 | 12 | 20 | 9 |
| Age75+(%) | 17 | 9 | 29 | 6 |
| White(%) | 72 | 75 | 65 | 69 |
| Black(%) | 14 | 10 | 19 | 19 |
| Other(%) | 14 | 15 | 16 | 12 |
| Lake(%) | 84 | 84 | 88 | 74 |
| Porter(%) | 16 | 16 | 12 | 26 |

We collect the preoperative PCR test time and results of patients in the hospital system, and demographic/geographic information including sex, age, race, and counties. As one of our study interests was to compare our analytic method to established symptomatic testing metrics, we collected the records for both asymptomatic presurgical and symptomatic patients tested within our hospital system, where the asymptomatic patients are assumed as our proxy sample to the target population. The symptomatic group is represented exclusively by outpatients tested with a positive answer to one or more queries about COVID 19 symptoms as defined by the Centers for Disease Control and Prevention (CDC); these queries have only changed over time in concert with changes made by CDC itself. Our data include daily records from Apr 28, 2020 to Feb 15, 2021, representing 30,116 asymptomatic and 13,960 symptomatic patients who received PCR tests. We poststratified the patients with tests to the 35,838 hospital EHR records in 2019 and the 654,890 community residents in Lake and Porter counties. Table 1 summarizes the test results and sociodemographic distributions, as well as the sociodemographics in the hospital system and the community, thus illustrating the discrepancy between the sample and the population.

The observed incidence rates are quite naturally different between the PCR tests: 1% for asymptomatic patients and 26% for symptomatic patients. As compared to the hospital system patients, asymptomatic patients with PCR tests tend to be female, middle-aged (35-64) or old (65-74), and white. For this reason, neither the hospital patients nor the asymptomatic patients serve as a precise representation of the community population, in particular with an under-coverage for young, male, and nonwhite residents. These differences are not large (Table 1); nonetheless, they are potential sources of error if not accounted for in our statistical model, and can also interfere with estimates of trends if the demographic breakdown of hospital patients varies over time. Furthermore, the county representation is unbalanced. Some patients are from south Cook County, Illinois, and are grouped into the Lake County as a proxy. Fortunately for our analysis, these contiguous communities have similar socioeconomic and ethnic demographics. The demographic discrepancy can be caused by unmeasured factors of the asymptomatic patients seeking elective surgeries, such as comorbidity status and healthcare utilization measures, the direct adjustment of which is impractical without their population

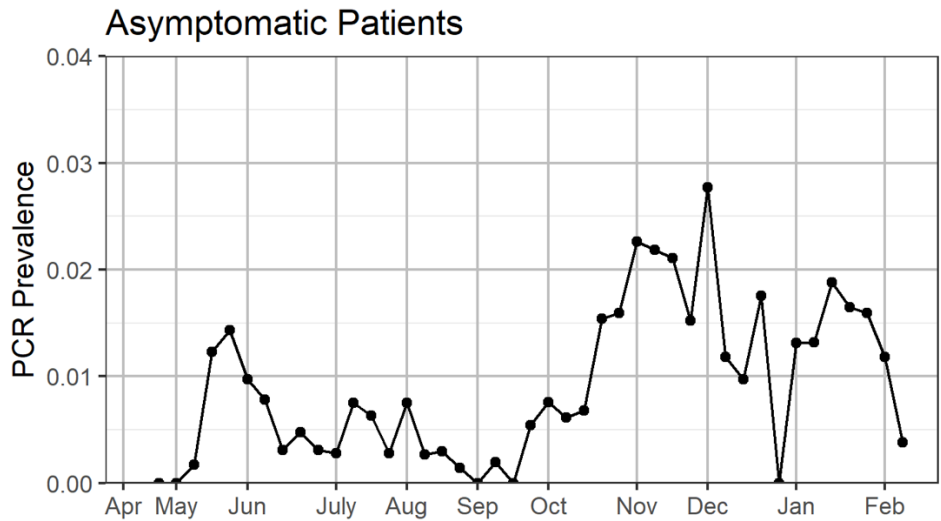

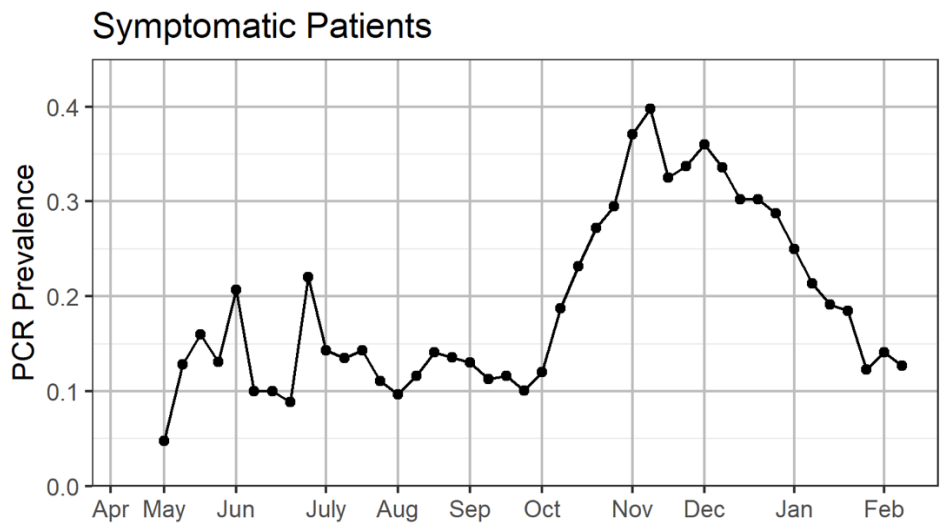

Figure 1. Observed weekly PCR test incidence for asymptomatic and symptomatic patients in the Community Hospital system. Note the different scales on the two graphs. The positions of the months on the x-axis correspond to the week of data containing the first of that month.

distribution. However, this confounding bias merely enhances the need to poststratify demographics.

Figure 1 presents the observed PCR test incidence over time for asymptomatic and symptomatic patients. The two groups present different prevalence magnitudes and trends. The prevalence changed over time with low values until September, then we see an increasing trend, with a spike at the end of October, a decrease in November and December, and a bounce back in January.

We present the weekly number of asymptomatic patients seeking elective surgeries in the hospital system, which shows stable sample sizes, and examine the observed sociodemographic distributions of asymptomatic and symptomatic patients receiving PCR tests over time and find that the asymptomatic patients' profiling is stable, while the sample decomposition of the symptomatic patients changes over time. Details are presented in eFigures 1-2 of the Supplement. This discrepancy provides supporting evidence for our pre-study hypothesis that we should treat the asymptomatic samples as a substantially better proxy sample of the target hospital or community population than the corresponding symptomatic data.

The variation of prevalence could be due to various sample decompositions across time, but variation in thresholds for testing symptomatic people over time and demographic is certainly a likely factor. Overall, our analysis here calls into question relying upon symptomatic data trends—as is currently the norm—in understanding the underlying true viral trends in the community and argues that asymptomatic testing is likely to be a superior proxy.

To correct for discrepancies between the sample demographics and those of the community at large, as necessitated by the above observations, we next apply MRP to model the incidence and poststratify to the hospital and community population for representative prevalence estimates. The outputs are given in Figure 2. For asymptomatic patients, the estimated positive PCR test prevalence is lower than the raw value after a spike between May 19 and May 25 and generally lower than 0.5% through September 28. These findings reflect a low observed clinical burden of COVID-19 in our community after the initial March-April outbreak; see District 1 hospitalization (13) in eFigure 3 in the Supplement. We observe an increasing trend in October and then decreasing throughout November, with MRP adjusted rates inflated significantly. The trend has spikes in December and January and decreases since mid-January.

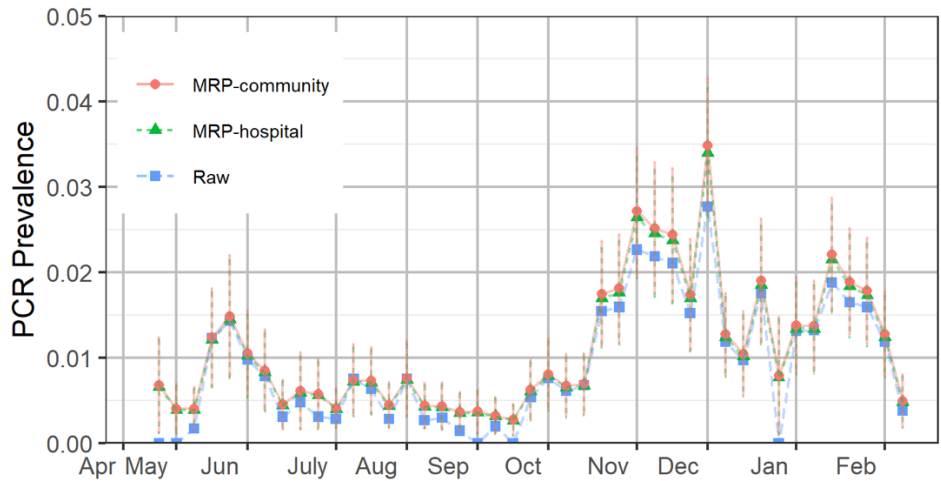

Figure 2: Estimated prevalence of the hospital system and community based on asymptomatic patients. The error bars represent one standard deviation of uncertainty. The positions of the months on the x-axis correspond to the week of data containing the first of that month.

Prediction metrics of clinical burden

We are interested in evaluating whether MRP-adjusted asymptomatic patients could track SARS-CoV-2 related hospitalization rates—as measured by counts of hospitalizations and emergency department (ED) visits—better than the currently applied metrics within our counties: positivity rate and counts of positive cases. Our expectation was that hospitalization census would lag viral incidence by a week or more and that ED COVID-19 visits would track actual viral incidence, perhaps with a few days' lag. These inferences follow from known lag times from exposure to symptoms to serious illness (12). To test this conjecture, we focus on the September 2020 through February 2021 interval as that timeframe encompasses all of the observed growth in viral burden after very low levels throughout late spring and the summer.

Our side-by-side analysis is illustrated in Figure 3. Each plot shows the week-to-week trend of the available metrics (asymptomatic MRP, positivity rate, and the number positive cases) and those of hospitalization rates (the number of hospitalizations and ED visits) within Lake and Porter Counties. Comparison with district and state data are in the Supplement.

All three metrics parallel hospitalization through September up until mid-October, after which the growth in positivity and counts of positive cases far outstrip the growth in hospitalization while the MRP data remains in strict parallel throughout. Further, the MRP estimates track even better with the ED visits. The hospitalization data, on the other hand, show a 1-week lag of the peak in November. Indeed, we begin to see some decrease in the MRP adjusted asymptomatic positives in November and December that parallels a decrease in hospitalization while the generally accepted state metrics continue to increase and even accelerate. These data suggest that ongoing increases in District 1 positive testing metrics may simply be artifacts of the test selection process, rather than actual growth in the viral spread. Overall, a comparison of trends shows that symptomatic positive cases only begin to decrease at the time hospitalization does, and fully a week after ED visits do. Positivity rate does not identify the apparent decrease in clinical burden and has in fact accelerated through that decrease.

DISCUSSION

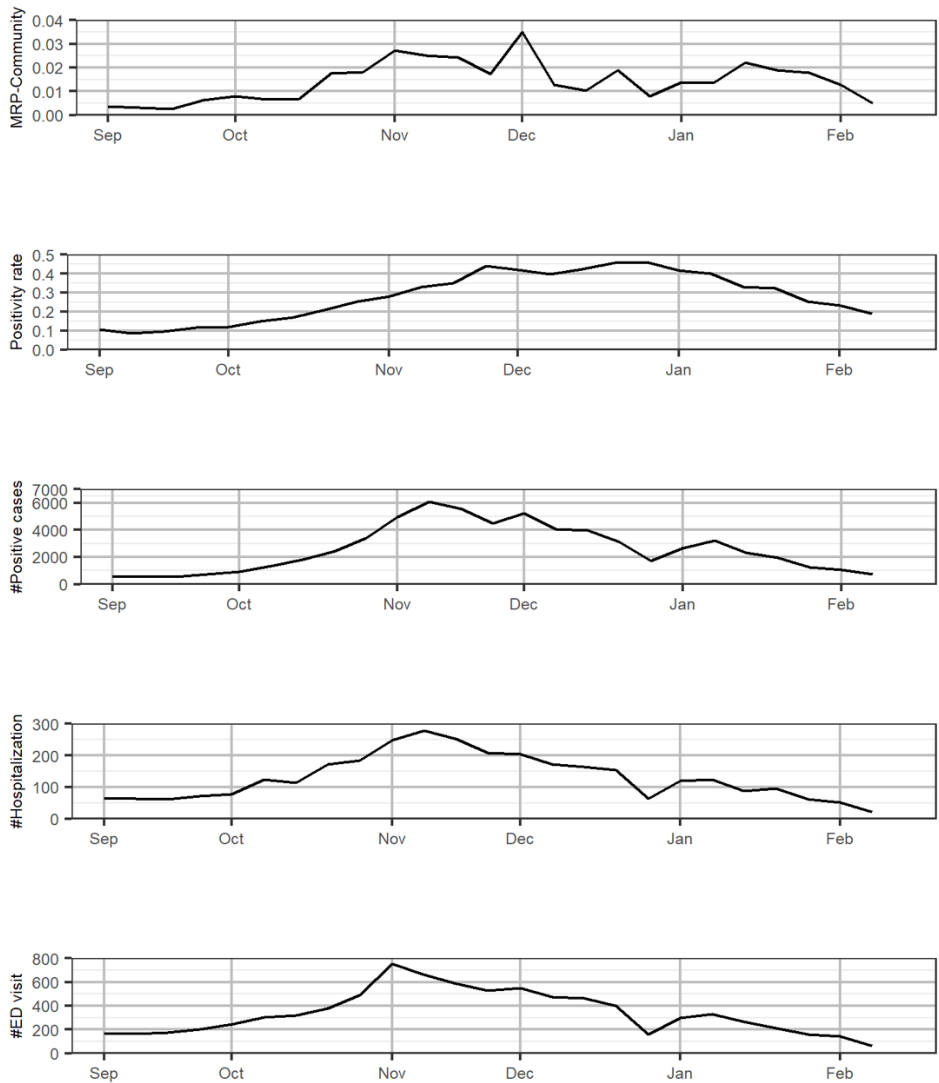

Figure 3: Comparison of MRP estimates with the reported hospitalization counts, ED visits, positivity rate, and the number of positive cases in Lake and Porter counties. The vertical dashed lines indicate the peak values. Note the different scales on the five graphs. The positions of the months on the x-axis correspond to the week of data containing the first of that month.

Our analysis indicates that applying our MRP normalization to data on the prevalence of asymptomatic SARS-CoV-2 infection produces a valuable leading indicator of hospital and community risk. When we set out to create a model for tracking viral incidence, we recognized substantial shortcomings in the available testing and its interpretation. While state data has become much richer and testing protocols more uniform since we started applying our model, selection bias is still a substantial concern. Our goal for this study was to develop an easily implemented testing strategy—available to any hospital system—that, after demographic and geographic adjustment, could reasonably approximate a representative sample. In so doing, we hoped to assess the reliability of currently accepted metrics in their prediction of virus trends and, if possible, to improve our ability to anticipate those trends.

The asymptomatic preoperative patients we identify with our protocol are a favorable group to build upon. All sizable hospital systems have a ready-made group of such patients who can produce a large number of data points quite rapidly. As patients continue to seek medical procedures, the population continues to naturally expand over time, and lends itself to trending data. In our nearly 900-bed hospital system, we have thus far generated over 30,000 data points over 43 weeks, representing a community of approximately 700,000 residents. The weekly number of data points has been fairly stable over time, and that observation is likely similar to many hospital systems. We have demonstrated that this sample population is fairly representative of the community demographics as a whole and that there has been minimal sample decomposition over time. That this population stability is not matched by similar demographic stability in the symptomatic population and that we are able to employ MRP to account for any demographic skew and instability in our own protocol both strongly argue that our model is far more representative of random sampling than the currently employed positive case and positivity data. We argue that hospital-based asymptomatic testing with MRP is a more reliably random metric than any currently available and is easily generated from the routine testing of patients prior to their scheduled procedures.

Having established a reasonable statistical validity for our model, we wished to use it to measure the reliability of current state-based metrics. Our analysis finds that in our community, all of the metrics trend similarly during viral surges. We would support the current view that numbers of positive cases and positivity both remain relatively stable during periods that our pseudorandom proxy method predicts to be stable and have increased during periods that our proxy predicts show true viral increases. Since the beginning of our study in early May, there have been significant changes in test availability and certainly anecdotal evidence that the indications for testing have changed quite a bit as well. Consequently, the number of tests and clinical indications for testing have almost certainly both increased considerably over that interval, but the patterns cited above have remained stable. For that reason, we feel that there are good reasons to believe that the validity of positive case counts and positivity as metrics for viral spread is, in the event, relatively insensitive to test numbers, test availability, and clinical thresholds for testing, at least in our community.

Finally, we wanted to test each of these metrics as predictors of clinical burden. During the entire study period, we have used our model to predict clinical needs: staffing, bed and ventilator availability, personal protection equipment supplies, and so forth. Our general observation was that this proxy provided us some useful lead time to prepare for the virus. When we compare our model's behavior to that of the standard metrics, we find it to be generally a better predictor of clinical burden. The effect is best seen in our November data. During the week of November 3-10, we were able to predict that viral transmission was decreasing and that our hospitalization was likely to be at or near its peak. Comparison of our model with ED COVID presentations in our area demonstrates a precise correlation. It is clear that new acute presentations correlate quite precisely with our metric, and that these changes occur about a week before positive cases and hospitalization census data change. Further, we see positivity rates continue to rise in our area well past the time that our metric and numbers of positive cases have declined. Given that ED visits and hospitalization census rates have also declined in that interval, we find that our model and the number of positive cases appear to be much better and more current predictors of the true viral clinical burden than positivity rates are.

In dealing with a case surge, the extra time of preparedness has been useful and nontrivial. Great benefit may also accrue from recognizing decreased transmission earlier, as we feel the model is able to do, as it may allow for opening up of needed clinical services and socioeconomic commerce in a community earlier than might otherwise be contemplated. In this sense, adherence to positivity rates may be particularly damaging.

We believe our model to be easily generalizable to many hospital systems. As discussed, the sample population and testing regime are readily available, likely to be reasonably representative and stable demographically over time, and easily normalized to true community demographics using the MRP code that we have made available. This approach represents a simple proxy for random sampling for any community that chooses to employ it. Further benefits might be gained by combining information from different hospital systems. The best way forward might be for individual hospitals and medical groups to gather and analyze their data as we propose in this article, with all the (de-identified) data shared in a common public repository, so that it would be possible for researchers to learn more by analyzing trends as they develop in the pooled dataset. This could be similar to other national data pooling efforts such as in the U.S. and Israel (14-15).

In addition, though, we demonstrate the clinical utility of less rigorous approaches as well. Should a system choose to track its patients according to our testing protocol, but not incorporate the MRP adjustments, the relative stability of the population demographics suggests that the trends remain quite valid. Our regression models have shown potentially strong effects of age and racial/ethnic status on our metric so that one would need to ensure at least reasonable stability of those particular traits to trust observed raw trends without formal MRP adjustment. We also find that while results depend strongly on the sensitivity of the test being employed, the trends in the results do not (details in the Supplement).

This finding is encouraging for longer-term monitoring. Very inexpensive antigen testing is now becoming broadly available. These tests may be less sensitive or more time-specific than the PCR-based RNA testing we have been using and therefore less able to verify the true magnitude of viral spread. Nonetheless, our data show that they will likely function perfectly well to follow viral transmission and clinical burden trends, especially if normalized by MRP. Practically speaking, these trends are the prime concern of most healthcare entities.

SUPPLEMENT

Here we present the supplemental materials on the summary of the paper structure, modeling details, and data showing demographic stability, comparison with publicly released records, and posterior predictive checks, for the paper, "Routine Hospital-based SARS-CoV-2 Testing Outperforms State-based Data in Predicting Clinical Burden."

Summary

Our primary goal is to track SARS-CoV-2 infection prevalence over time, and to do so in a reliable way. We are assuming that the ratio of the number of asymptomatic patients to that of the symptomatic patients is fixed. We would then need to normalize each group from the sample demographics to true demographics via MRP. Once that is accomplished, we posit that our metric represents an appropriate approximation of a true random sample. In particular, it either is superior to what is being done now (tracking positivity and overall positive case numbers) or it provides a respectable metric to verify that those currently used data are reasonable approximations of fact. All of our data representations should either support one of those contentions, or provide evidence supporting or refuting critiques of our approach. To that end, we need to follow prevalence changes over time, but verify (through MRP and statistical analysis) that these trends are real changes in the community and not changes in the sample demographics. We then compare the MRP normalized trends in our model with currently employed metrics of viral prevalence trends: namely, positivity rates and numbers of positive tests in the community. Finally, we compare these metrics with hospitalization rates to determine how predictive our model and the current metrics may be of the community clinical burden of the virus.

Modeling details

We denote the test result for individual i as $y_i$, where $y_i = 1$ indicating a positive result and $y_i = 0$ indicating negative. Let $p_i = \Pr(y_i = 1)$ be the probability that this person tests positive. The analytic incidence $p_i$ is a function of the sensitivity $\delta$, specificity $\gamma$, and the true viral incidence $\pi_i$ for individual i: $p_i = (1-\gamma)(1-\pi_i) + \delta\pi_i$. We fit a logistic regression for $\pi_i$ with covariates including sex, age, race, county, time, and the two-way interaction between sex and age. Using the model-predicted incidence $\hat{\pi}_i$, we apply the sociodemographic distributions in the hospital system and the community to generate the population level prevalence estimates, as the poststratification step in MRP.

We use the following logistic regression in (1) to allow time variation of prevalence over time in the multilevel model parameters.

$$\text{logit}(\pi_i) = \beta_1 + \beta_2 \text{male}_i + \alpha^{\text{age}}_{\text{age}[i]} + \alpha^{\text{race}}_{\text{race}[i]} + \alpha^{\text{county}}_{\text{county}[i]} + \alpha^{\text{time}}_{\text{time}[i]} + \alpha^{\text{age}*\text{male}}_{\text{age}*\text{male}[i]},$$

where $\text{male}_i$ is an indicator taking on the value 0.5 for men and -0.5 for women; age[i], race[i], and county[i] represent age, race, and county categories, with a two-way interaction term age * male[i]; time[i] indices the time in weeks when the test result is observed for individual i; and the α parameters are vectors of varying intercepts to which we assign hierarchical priors:

$$\alpha^{\text{name}} \sim \text{normal}(0, \sigma^{\text{name}}), \quad \sigma^{\text{name}} \sim \text{normal}_+(0, 2.5),$$

for name ∈ {age, race, county, age ∗ male}. Here, $\text{normal}_+(a, b)$ represents a half-normal distribution with the mean $a$ and standard deviation $b$ restricted to positive values. And we set the time-varying effect: $\alpha^{\text{time}} \sim \text{normal}(0, \sigma^{\text{time}})$, $\sigma^{\text{time}} \sim \text{normal}_+(0, 5)$, to allow for the possibility of large variations across time. The larger the estimated variation, the larger effects of the predictors. Assume the prior information for the unknown sensitivity $\delta$ and specificity $\gamma$ includes: $y_\gamma$ negative results in $n_\gamma$ tests of known negative subjects and $y_\alpha$ positive results from $n_\alpha$ tests of known positive subjects. The model for the number of positive results y out of n tests is specified as

$$y_\gamma \sim \text{binomial}(n_\gamma, \gamma), \quad y_\delta \sim \text{binomial}(n_\delta, \delta).$$

According to the test protocol, the sensitivity is around 70%, and the specificity is around 100%. We solicit prior information from previous testing results (2). For the sensitivity, the prior data $y_\delta/n_\delta$ are 70/100, 78/85, 27/37, and 25/35; and the prior data for the specificity $y_\gamma/n_\gamma$ are 0/0, 368/371, 30/30, 70/70, 1102/1102, 300/300, 311/311, 500/500, 198/200, 99/99, 29/31, 146/150, 105/108, and 50/52.

After fitting the Bayesian model, we adjust for the selection bias by applying the sociodemographic distributions in the hospital system and the community to generate the population level prevalence estimates, as the poststratification step in MRP. For each of the 2 ∗ 5 ∗ 3 ∗ 2 cells in the cross-tabulation table of sex (2 levels), age (5 levels), race (3 levels) and county (2 levels), we have the cell-wise incidence estimate $\hat{\pi}_j$, and population count $N_j$, where j is the cell index, and calculate the weekly prevalence estimate in the population,

$$\pi_{\text{avg}} = \sum_j N_j \hat{\pi}_j / \sum_j N_j.$$

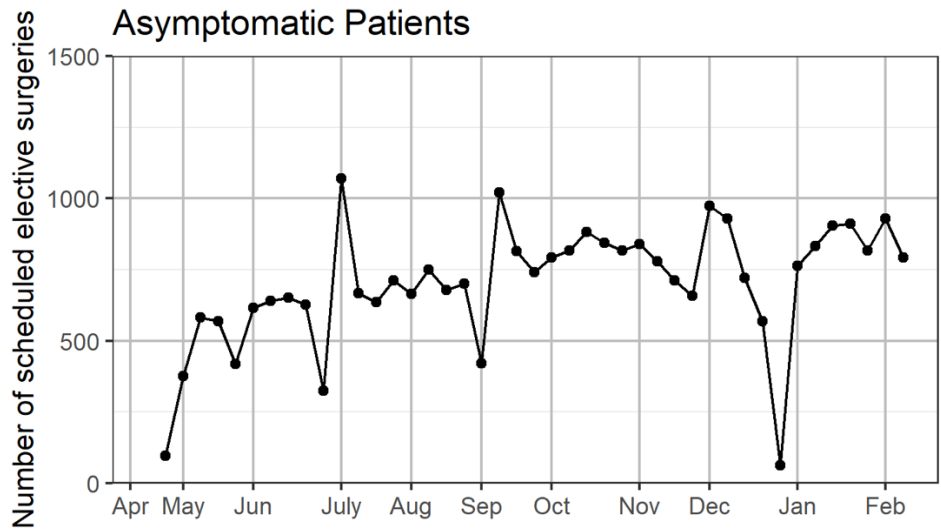

eFigure 1: Number of elective surgeries for asymptomatic patients each week in the Community Hospital system. The positions of the months on the x-axis correspond to the week of data containing the first of that month.

Demographic stability

We examine the weekly number of asymptomatic patients seeking elective procedures, the sample sizes of asymptomatic patients, and the frequency distributions of asymptomatic and symptomatic patients across week in different sociodemographic (sex, race, and age) groups.

eTable1. Weekly sample sizes available across time in different groups (week 18: 4/28/2020 - 5/3/2020, week 60: 2/8/2021 - 2/14/2021).

| Week | Total | Male | Female | Age0-17 | Age18-34 | Age35-64 | Age65-74 | Age75+ | White | Black | Other |
|---|---|---|---|---|---|---|---|---|---|---|---|
| 18 | 95 | 47 | 48 | 1 | 12 | 42 | 26 | 14 | 65 | 16 | 14 |
| 19 | 376 | 173 | 203 | 17 | 40 | 176 | 91 | 52 | 274 | 55 | 47 |
| 20 | 582 | 242 | 340 | 26 | 73 | 257 | 123 | 103 | 403 | 83 | 96 |
| 21 | 569 | 216 | 353 | 26 | 81 | 246 | 130 | 86 | 415 | 73 | 81 |
| 22 | 419 | 164 | 255 | 16 | 53 | 171 | 105 | 74 | 302 | 66 | 51 |
| 23 | 616 | 244 | 372 | 26 | 75 | 259 | 150 | 106 | 451 | 95 | 70 |
| 24 | 640 | 271 | 369 | 24 | 46 | 304 | 155 | 111 | 476 | 81 | 83 |
| 25 | 651 | 270 | 381 | 26 | 66 | 292 | 166 | 101 | 476 | 79 | 96 |
| 26 | 627 | 288 | 339 | 24 | 66 | 270 | 151 | 116 | 457 | 86 | 84 |
| 27 | 324 | 136 | 188 | 9 | 39 | 141 | 62 | 73 | 249 | 37 | 38 |
| 28 | 1070 | 435 | 635 | 46 | 114 | 466 | 248 | 196 | 793 | 148 | 129 |
| 29 | 668 | 273 | 395 | 17 | 66 | 306 | 173 | 106 | 471 | 94 | 103 |
| 30 | 635 | 266 | 369 | 17 | 70 | 279 | 145 | 124 | 472 | 74 | 89 |
| 31 | 711 | 292 | 419 | 29 | 72 | 326 | 149 | 135 | 490 | 103 | 118 |
| 32 | 665 | 267 | 398 | 20 | 81 | 287 | 163 | 114 | 487 | 86 | 92 |
| 33 | 750 | 312 | 438 | 18 | 101 | 321 | 185 | 125 | 527 | 115 | 108 |
| 34 | 678 | 262 | 416 | 10 | 85 | 330 | 157 | 96 | 477 | 91 | 110 |
| 35 | 700 | 266 | 434 | 17 | 74 | 313 | 171 | 125 | 494 | 96 | 110 |
| 36 | 420 | 163 | 257 | 13 | 60 | 181 | 92 | 74 | 293 | 56 | 71 |
| 37 | 1021 | 380 | 641 | 35 | 111 | 444 | 232 | 199 | 721 | 155 | 145 |
| 38 | 815 | 325 | 490 | 23 | 60 | 389 | 205 | 138 | 583 | 129 | 103 |
| 39 | 742 | 324 | 418 | 19 | 67 | 350 | 170 | 136 | 548 | 103 | 91 |
| 40 | 792 | 316 | 476 | 19 | 75 | 364 | 210 | 124 | 573 | 111 | 108 |
| 41 | 817 | 307 | 510 | 18 | 73 | 378 | 189 | 159 | 599 | 113 | 105 |
| 42 | 883 | 341 | 542 | 23 | 94 | 416 | 203 | 147 | 639 | 126 | 118 |
| 43 | 843 | 348 | 495 | 32 | 82 | 389 | 189 | 151 | 583 | 131 | 129 |
| 44 | 816 | 341 | 475 | 18 | 84 | 381 | 202 | 131 | 592 | 114 | 110 |
| 45 | 839 | 337 | 502 | 28 | 74 | 399 | 206 | 132 | 598 | 117 | 124 |
| 46 | 778 | 315 | 463 | 23 | 69 | 368 | 182 | 136 | 562 | 107 | 109 |
| 47 | 712 | 294 | 418 | 15 | 77 | 352 | 149 | 119 | 507 | 104 | 101 |
| 48 | 658 | 278 | 380 | 10 | 70 | 298 | 167 | 113 | 495 | 81 | 82 |
| 49 | 975 | 406 | 569 | 24 | 97 | 482 | 232 | 140 | 719 | 128 | 128 |
| 50 | 930 | 385 | 545 | 36 | 109 | 427 | 223 | 135 | 678 | 118 | 134 |
| 51 | 720 | 297 | 423 | 23 | 79 | 345 | 174 | 99 | 513 | 107 | 100 |
| 52 | 569 | 218 | 351 | 13 | 68 | 292 | 129 | 67 | 427 | 60 | 82 |
| 53 | 62 | 16 | 46 | 2 | 14 | 31 | 9 | 6 | 42 | 14 | 6 |
| 54 | 763 | 308 | 455 | 9 | 86 | 364 | 175 | 129 | 553 | 114 | 96 |
| 55 | 833 | 346 | 487 | 21 | 97 | 371 | 188 | 156 | 595 | 122 | 116 |
| 56 | 904 | 387 | 517 | 21 | 77 | 431 | 220 | 155 | 662 | 122 | 120 |
| 57 | 911 | 389 | 522 | 15 | 72 | 441 | 244 | 139 | 636 | 154 | 121 |
| 58 | 816 | 329 | 487 | 25 | 59 | 404 | 205 | 123 | 584 | 119 | 113 |

| 59 | 929 | 399 | 530 | 32 | 86 | 447 | 207 | 157 | 680 | 132 | 117 |
| 60 | 792 | 330 | 462 | 23 | 70 | 354 | 223 | 122 | 580 | 91 | 121 |

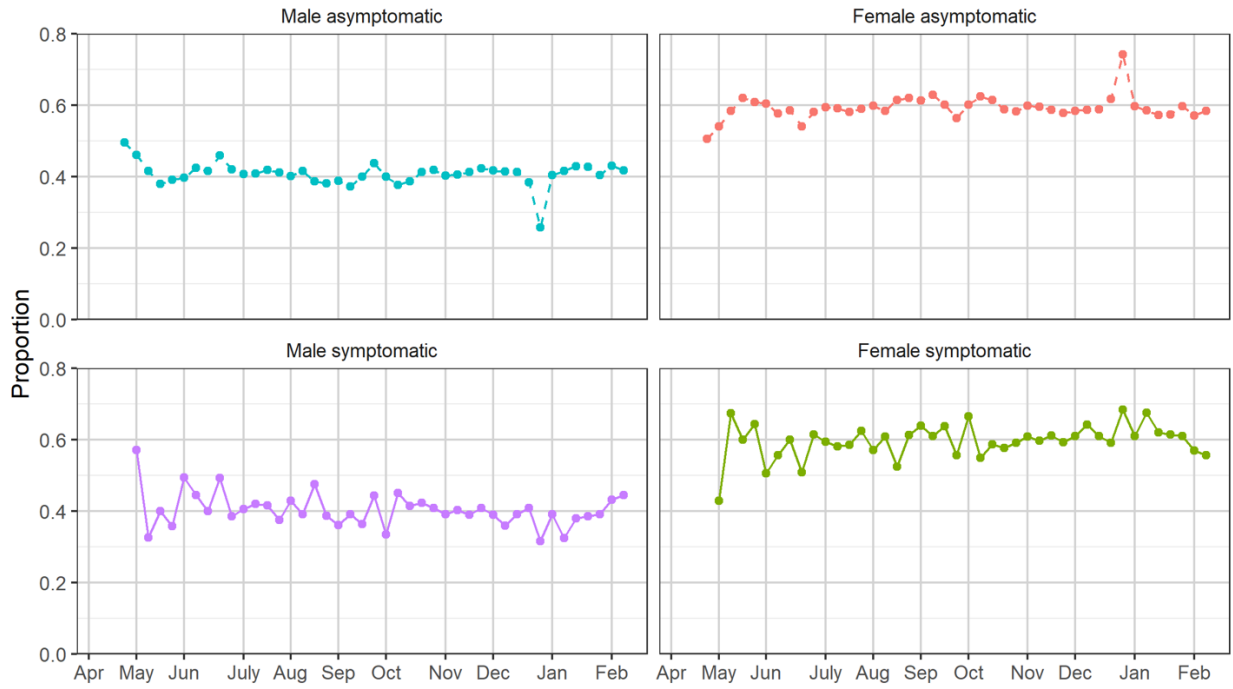

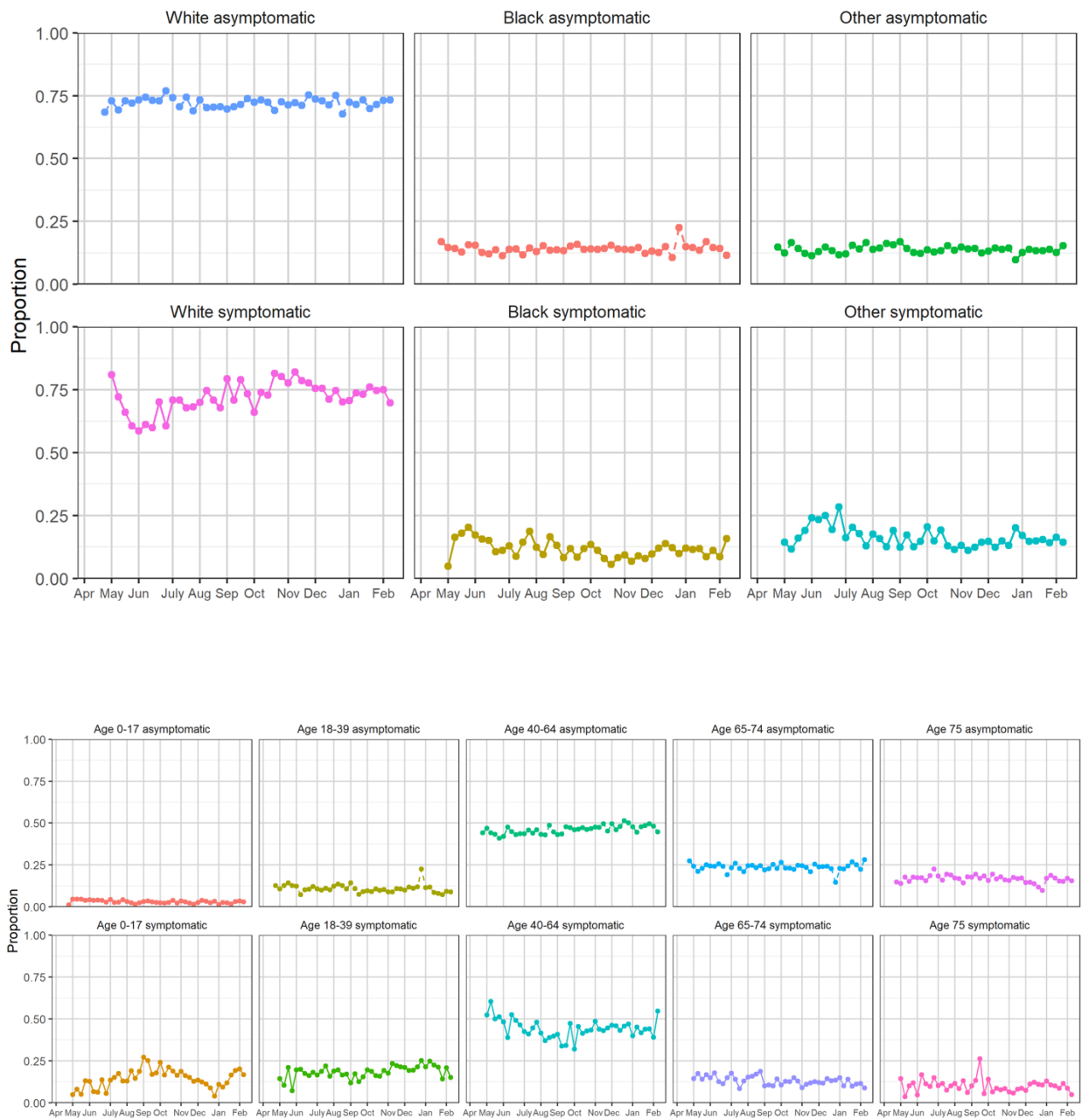

eFigure 2: Demographic distributions of asymptomatic and symptomatic patients across time. The positions of the months on the x-axis correspond to the week of data containing the first of that month.

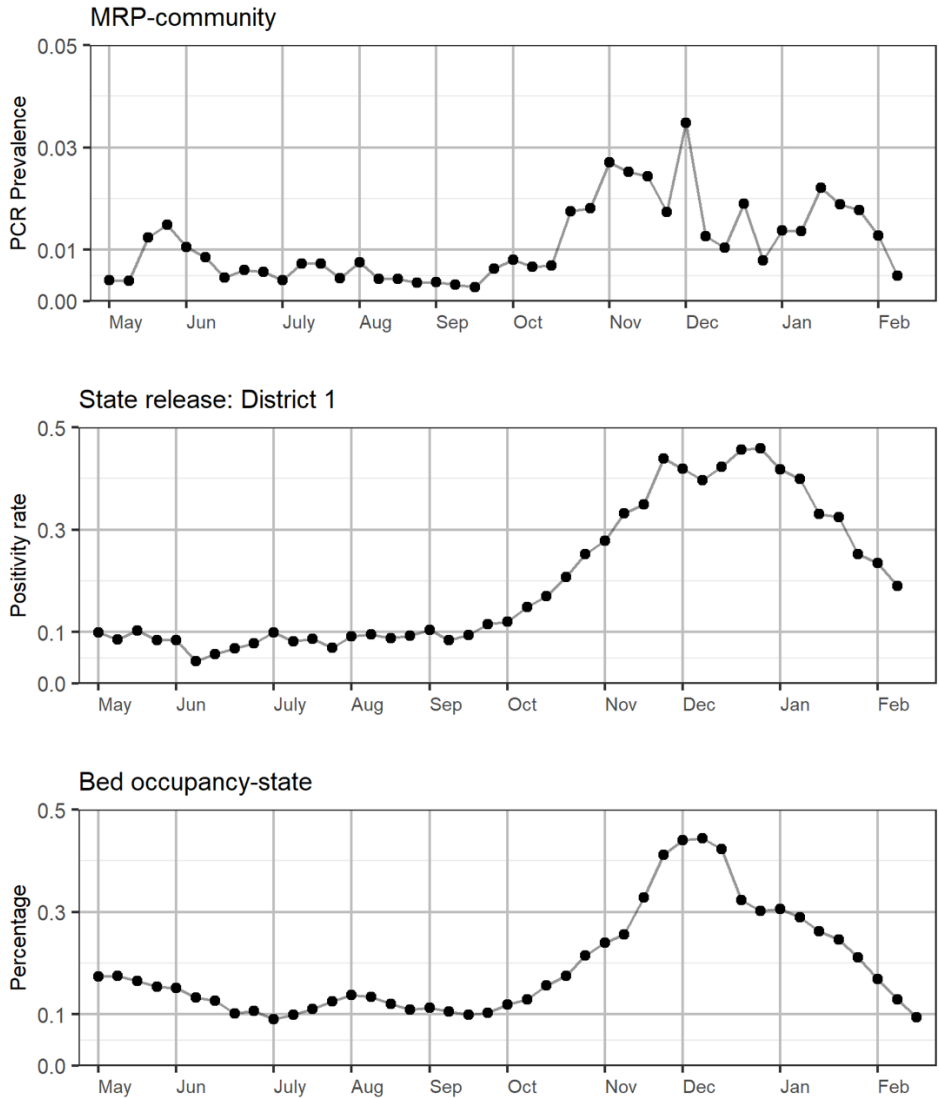

eFigure 3: Comparison of MRP estimates with reported test positivity rate in District 1 and hospital bed occupancy rates in the state, Indiana. Note the different scales on the three graphs. The positions of the months on the x-axis correspond to the week of data containing the first of that month.

Comparison with publicly released data

eFigure 3 compares the MRP estimates of asymptomatic patients with the publicly released positivity rate data in District 1, where our hospital system is located, and the state hospital bed occupancy rate. The district-level test positivity rate is higher than that of MRP estimates, as expected, since the state data largely involves the testing of symptomatic patients. The two positivity rates both have an increasing trend since September, but deviate in November, where MRP estimates start decreasing. The occupancy rate of hospital beds in the data generally follows a similar trend but presents a lower increasing rate in November than the positivity rate in District 1. We observe that MRP asymptomatic data are able to predict clinical behavior a week or two earlier than the District 1 positivity data.

Posterior predictive check

To evaluate the model fitting, we apply a posterior predictive check by generating replicated data from the posterior model distributions with the same sample size as the raw data. We use the collected sample decomposition records every week and estimated prevalence rates of poststratification cells, defined by the cross-tabulation of age, gender, race/ethnicity, and county information, to generate replicated test results. We compare the weekly prevalence rates between the replicated data and observed data. eFigure 4 shows that the model of asymptomatic patients can capture the raw data structure, implying that this aspect of the data is captured well by the fitted model.

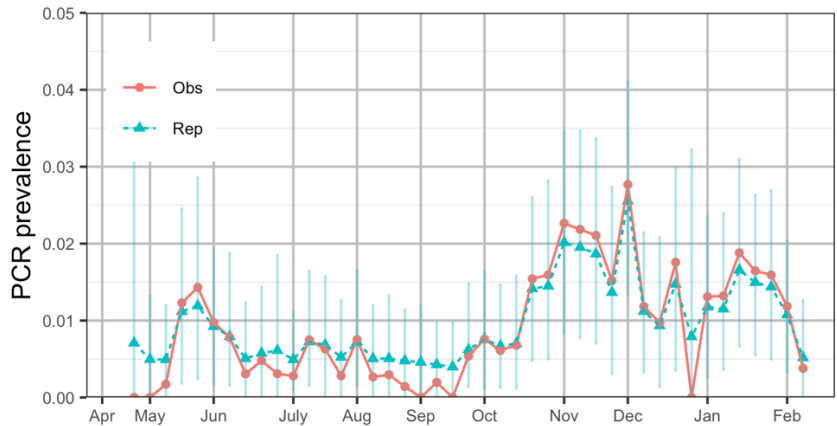

eFigure 4: Posterior predictive check: comparison of replicated and observed prevalence. The error bars represent 95% predictive intervals. The positions of the months on the x-axis correspond to the week of data containing the first of that month.

We have performed extensive sensitivity analyses of the estimates by changing the modeling mean structure and prior specifications, for example, using spline functions of time, assigning a flexible Gaussian process regression model as the prior distribution of time-varying effects, and changing hyperparameter values. The findings of the hospital- and community-level prevalence estimates are robust without changing conclusions.

We account for the uncertainty of sensitivity and specificity in a Bayesian framework and use the meta-analysis study findings as the prior specification (1). The presented results above are based on the prior information concentrated on the sensitivity value of 70% and specificity 100%. When we set the prior sensitivity data as 70/100, the MRP estimates are similar under the current prior setting. We also compare the PCR results when the prior sensitivity value is set at 65%, 60% and 55%; the results are approximately inflated by the reciprocal of the value of sensitivity, suggesting that the magnitude of our estimates is sensitive to the quality of PCR tests. Nevertheless, the trends within a given sensitivity remain stable.